\documentclass[apex]{jjap3}
\usepackage{txfonts}
\usepackage{cite}

\title{Pressure Effect in Bi-2212 and Bi-2223 Cuprate Superconductor}
\author{Shintaro ADACHI$^1$\thanks{ADACHI.Shintaro@nims.go.jp}, Ryo MATSUMOTO$^{1, 2}$, Hiroshi HARA$^{1,2}$, Yoshito SAITO$^{1, 2}$, Peng SONG$^{1, 2}$, Hiroyuki TAKEYA$^1$, Takao WATANABE$^3$, and Yoshihiko TAKANO$^{1, 2}$}
\inst{$^1$MANA, National Institute for Materials Science (NIMS), 1-2-1 Sengen, Tsukuba, Ibaraki 305-0047, Japan \\
$^2$Graduate School of Pure and Applied Sciences, University of Tsukuba, 1-1-1 Tennodai, Tsukuba, Ibaraki 305-8577, Japan \\
$^3$Graduate School of Science and Technology, Hirosaki University, 3 Bunkyo-cho, Hirosaki, Aomori 036-8561, Japan}
\abst{We report the pressure effect in Bi$_2$Sr$_2$Ca$_2$Cu$_3$O$_{10+{\delta}}$ (Bi-2223) single crystal with a small amount of intergrowth of Bi$_2$Sr$_2$CaCu$_2$O$_{8+{\delta}}$ (Bi-2212). Their superconducting transition temperatures $\it T$$_c$s showed a dome-like shape as a function of pressure, which showed a good agreement with the general relation between the carrier concentration and $\it T$$_c$. Our experimental results indicate that high pressure can induce effective carrier doping into the multilayered high-$\it T$$_c$ cuprate superconductor.}

\begin{document}
\maketitle

Several studies reported that the superconducting transition temperature ($\it T$$_c$) of cuprate superconductors can be enhanced by applying pressure\cite{Gao1994, Monteverde2005, Takeshita2013, Mito2017, Mito2016, Murayama1991, Yoneda1990, Watanabe2003, Chen2010}. The highest $\it T$$_c$ above 160 K among cuprates has been observed in HgBa$_2$Ca$_2$Cu$_3$O$_{8+{\delta}}$ (Hg-1223) under 20-30 GPa\cite{Gao1994, Monteverde2005, Takeshita2013}. According to recent results investigating the pressure effect up to about 20 GPa in Hg-1223\cite{Mito2017} and YBa$_2$Cu$_4$O$_8$ (Y-124)\cite{Mito2016}, $\it T$$_c$ shows a dome-like shape as a function of applied pressure $\it P$, which is similar to the general relationship between the carrier concentration and $\it T$$_c$. The pressure dependence of the Hall coefficient in YBa$_2$Cu$_3$O$_{7-{\delta}}$, Y-124, Bi$_2$Sr$_2$CaCu$_2$O$_{8+{\delta}}$ and Tl$_2$Ba$_2$CuO$_{6+{\delta}}$, also indicates that effective carrier doping can be induced by pressure\cite{Murayama1991}. The structure of cuprates alternately stacks a superconducting CuO$_2$ plane and insulating charge reservoir layer (CRL). It is empirically known that the $\it T$$_c$ of cuprate superconductors enhances with increasing the number of CuO$_2$ planes ($\it n$) per unit cell from 1 to 3 and decreases with further increasing $\it n$ from 4\cite{Maeda1988, Akimitsu1987, Adachi1988, Parkin1988, Schilling1993, Scott1994, Karppinen1999, Iyo2001}. It should be noted that the number of CRLs does not change even when $\it n$ increases. The homologous series of the high-$\it T$$_c$ cuprates are described by two universal composition ratios, M-12($\it n$-1)$\it n$ or M-22($\it n$-1)$\it n$, where M is an element forming the CRL. As the distance between the CRLs and inner CuO$_2$ planes increases, the doping level of inner CuO$_2$ planes decreases\cite{Karppinen1999}. The nuclear magnetic resonance measurements in the multilayered cuprates\cite{Tokunaga1999, Kotegawa1999, Mukuda2006, Shimizu2011, Mukuda2012, Iwai2014} showed that the carrier density of outer CuO$_2$ planes is always larger than that of  inner CuO$_2$ plane. Consequently, the average of the carrier concentration in CuO$_2$ plane decreases as $\it n$ increases. Chemical doping to the multilayered cuprate is extremely difficult, however, high pressure may induce effective doping. Up to now, few studies have focused on the relationship between pressure effect of $\it T$$_c$ and the number of CuO$_2$ planes $\it n$. To investigate this relation, it is necessary to directly observe the $\it T$$_c$ under high pressure in each cuprates having different $\it n$.
\par In this study, we performed the resistance measurements under high pressure up to 24.4 GPa using triple-layered Bi$_2$Sr$_2$Ca$_2$Cu$_3$O$_{10+{\delta}}$ (Bi-2223) single crystal which contains a small amount of double-layered Bi-2212 as an intergrowth structure. The purpose of the study was to investigate the relationship between $\it P$-$\it T$$_c$ curves and the number of CuO$_2$ planes $\it n$ in Bi-based cuprate superconductor\cite{Maeda1988}. Both Bi-2223 and intergrowth of Bi-2212 have comparable CRLs, and Bi-2223 has one more the CuO$_2$ planes than Bi-2212. Comparison of $\it T$$_c$ of Bi-2223 and Bi-2212 is suitable in order to investigate the pressure effect of $\it T$$_c$ as a function of $\it n$ purely. In addition, Bi-22($\it n$-1)$\it n$ with two CRLs possibly indicates the pressure-induced carrier doping larger than such Hg-12($\it n$-1)$\it n$ with one CRL\cite{Gao1994, Monteverde2005, Takeshita2013, Mito2017}.
The single crystals of Bi-2223 were grown using the traveling solvent floating zone (TSFZ) method as reported in Ref. 12. The samples were prepared from powders of Bi$_2$O$_3$, SrCO$_3$, CaCO$_3$, and CuO (all of 99.9$\%$ purity or higher) with the cation ratio of Bi : Sr : Ca : Cu = 2.2 : 1.9 : 2.0 : 3.0. The doping level of obtained Bi-2223 is slightly under-doped state\cite{Adachi2015-2, Fujii2001, Iye2010, Adachi2015}.

The Bi-2223 single crystals were characterized using a 2$\theta$/$\theta$ scan of X-ray diffraction (XRD) after the $\theta$ scan. The magnetic susceptibility was measured for cleaved Bi-2223 single crystal under 10 Oe using the magnetic property measurement system (MPMS, Quantum Design). The $\it R$-$\it T$ measurements under high pressure of Bi-2223 single crystals containing intergrowth of the Bi-2212 were performed using an originally designed diamond anvil cell (DAC)\cite{Matsumoto2016, Matsumoto2018, Matsumoto2018-2}. The pressures in this study were measured at room temperature. As shown in the previous study\cite{Matsumoto2016}, the room temperature pressure was in good agreement with the low temperature one in our DAC configuration.

The crystal structure of Bi-2212 and Bi-2223 are shown in figure 1. These are typical layered materials as the homologous series in the high-$\it T$$_c$ cuprate. 

Figure 2a shows a XRD pattern for the Bi-2223 single crystal. All of the peaks were assigned to the (0 0 2$\it l$) peaks of Bi-2223. Figure 2b is the enlargement of XRD data around the (0 0 10) peak, which is as sharp as FWHM=0.173 deg. Intergrowth of Bi-2212 was not detected using the XRD analysis. The above fact indicates that single crystal is well developed perpendicular to the ab plane.

\par Figure 3a and 3b show the temperature dependence of magnetic susceptibility for the Bi-2223 single crystal. The superconducting transition was observed at 105 K when the magnetic field was applied perpendicular to the ab plane of Bi-2223. On the other hand, two superconducting transitions were observed at 83 K and 105 K, when the magnetic field was applied parallel to the ab plane. The Bi-2212 intergrows lamellarly in Bi-2223 crystals\cite{Fujii2001}. Therefore, magnetic field parallel to the ab plane can penetrate into the normal state Bi-2212 above 83 K.  On the other hand, magnetic field perpendicular to the ab plane hardly penetrates into the Bi-2212 lamellae because, in this configuration, the small amounts of Bi-2212 are sheltered from the magnetic field by Bi-2223. Observation of two $\it T$$_c$s of both Bi-2223 and Bi-2212 using such a crystal is an advantage in $\it R$-$\it T$ measurements.

\par Figure 4a and 4b show the temperature dependence of the resistance under various pressures, and figure 4c is the enlarged view of a sample space with a Bi-2223 single crystal mounted in the DAC. There were two resistive drops caused by the superconducting transitions of Bi-2223 and intergrowth of Bi-2212, corresponding to $\it T$$_{c1}$ (higher) and $\it T$$_{c2}$ (lower) defined in figure 4a, respectively. After releasing the pressure, $\it T$$_{c1}$ and $\it T$$_{c2}$ were 105.4 K and 84.5 K respectively, indicating a good agreement with $\it T$$_c$s before applying pressure. Temperature dependence of the normal state resistance showed semiconducting behavior with increasing pressure. This behavior was irreversible as a function of applying pressure since those semiconducting behaviors maintained even after releasing the pressure.  We suppose that the samples have been damaged under high pressures. 
Figure 5a shows a plot of pressure ($\it P$) versus $\it T$$_c$s. We found that both $\it T$$_c$s increased initially with an increasing pressure, reached a maximum at each specific pressure value, and then decreased with further pressure. Moreover, Bi-2223 shows the $\it T$$_c^{MAX}$ at lower pressure than Hg-1223 by comparison of pressure dependence of $\it T$$_c$ of Bi-2223 ($\it T$$_{c1}$) and that of Hg-1223 in previous studies\cite{Gao1994, Monteverde2005, Takeshita2013, Mito2017}. We believe that Bi-2223 having two CRLs is effectively pressure-induced doped compared to Hg-1223 as assumed in the introduction. 
On the other hand, Figure 5b quantitatively compares the $\it T$$_c$s versus normalized pressure ($\it P$ / $\it n$), where the transverse axis is the value of each pressure divided by the number of CuO$_2$ planes $\it n$ (=3) in Bi-2223 to that (=2) in Bi-2212 respectively, and longitudinal axis is $\it T$$_c$s normalized by each $\it T$$_c^{MAX}$. As a result, both normalized $\it P$-$\it T$$_c$ curves almost agree with each other.
The pressure dependence of $\it T$$_c$ for Bi-2223 and Bi-2212 show dome-like behavior suggesting that effective carrier doping can be induced by pressure as well as previous works\cite{Mito2017, Mito2016, Murayama1991}. The average of the carrier concentration in CuO$_2$ plane decreases as $\it n$ increases. Consequently, Bi-2223 required a larger pressure than Bi-2212 in order to show the optimum $\it T$$_c$. We have demonstrated that $\it P$/$\it n$-$\it T$$_c$ curves show a good agreement with the general relation between the carrier concentration and $\it T$$_c$. This indicates that $\it T$$_c$ can be enhanced by applying high pressure to introduce carrier to the CuO$_2$ plane from CRL even when the number of CuO$_2$ planes $\it n$ increases. Therefore, if the pressure can be applied to Bi-2234 (n=4) or more multilayered BSCCO, it would show further high-$\it T$$_c$ than Bi-2223 at higher pressure.

In summary, the $\it R$-$\it T$ measurements were carried out under high pressure using a Bi-2223 single crystal with a small amount of intergrowth of Bi-2212. We have succeeded in simultaneously observing the pressure effect of $\it T$$_c$ in both Bi-2223 and Bi-2212. The obtained $\it P$/$\it n$-$\it T$$_c$ curves show dome-like behavior, which shows a good agreement with the general relation as the carrier concentration versus $\it T$$_c$. Therefore, the high pressure can introduce carrier to the CuO$_2$ plane in the multilayered high-$\it T$$_c$ cuprate superconductor. More details of the pressure effect in multilayered high-$\it T$$_c$ cuprate superconductors remain of an important issue for the future research.

\begin{figure}
\includegraphics{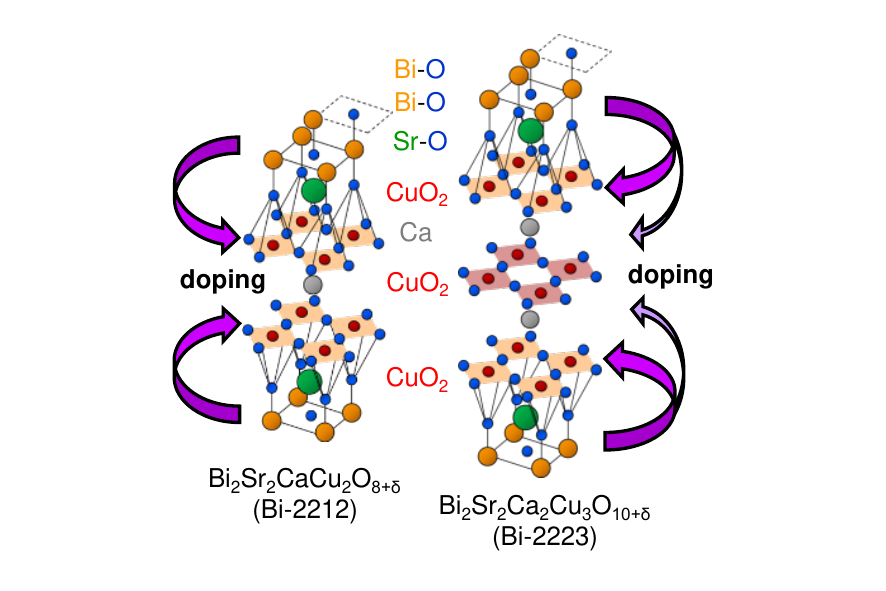}
\caption{(color online). Crystal structure of double-layered Bi-2212 and triple-layered Bi-2223. The thickness of the arrows show the schematic image of the doping strength.}
\label{f1}
\end{figure}

\begin{figure}
\includegraphics{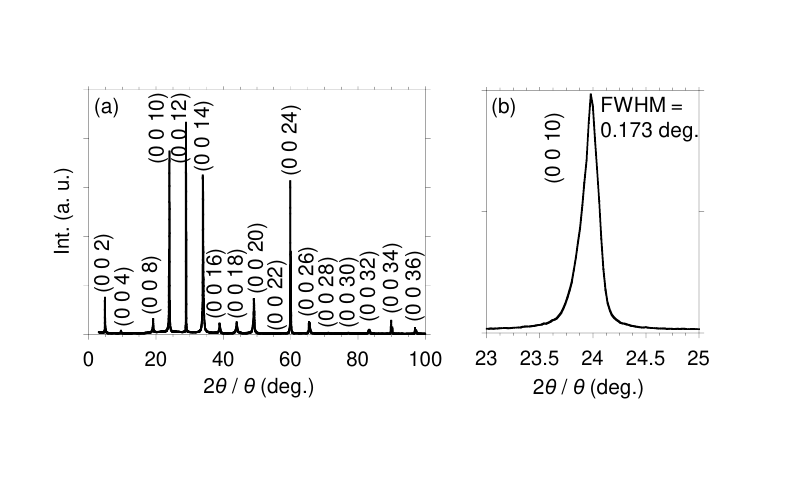}
\caption{(a) Typical XRD patterns using CuK$\alpha$ radiation for obtained Bi-2223 single crystal. (b) Enlargement of around a (0 0 10) peak.}
\label{f2}
\end{figure}

\begin{figure}
\includegraphics{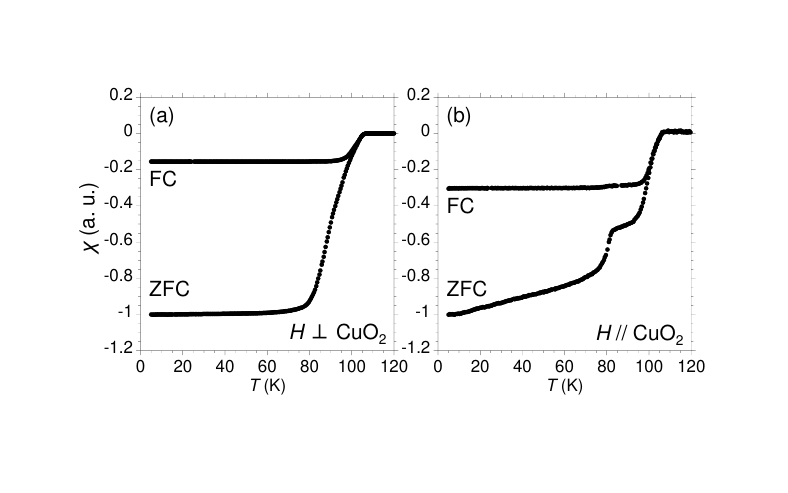}
\caption{Temperature dependence of magnetic susceptibility for obtained Bi-2223 single crystal. (a) Magnetic fields (= 10 Oe) were applied perpendicularly to the CuO$_2$ planes, and (b) parallel to the CuO$_2$ planes using a same single crystal in both data.}
\label{f3}
\end{figure}
 
\begin{figure}
\includegraphics{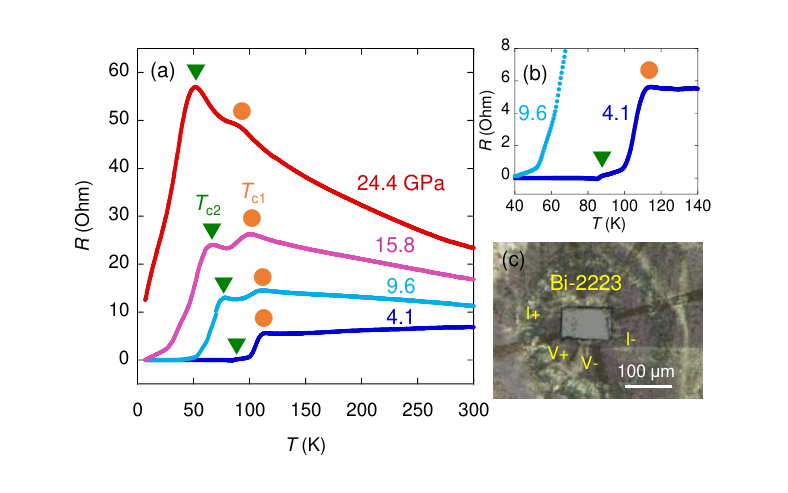}
\caption{(color online). (a) Temperature dependence of resistance under various pressures for a Bi-2223 single crystal with little intergrowth of Bi-2212. (b) Enlarged view of $\it RT$ curve under 4.1 GPa around $\it T$$_c$s. The dot circles show the higher superconducting transition temperature ($\it T$$_{c1}$). The inverse triangle dots show lower $\it T$$_c$ ($\it T$$_{c2}$). (c) Enlargement of sample space with a Bi-2223 single crystal.}
\label{f4}
\end{figure}

\begin{figure}
\includegraphics{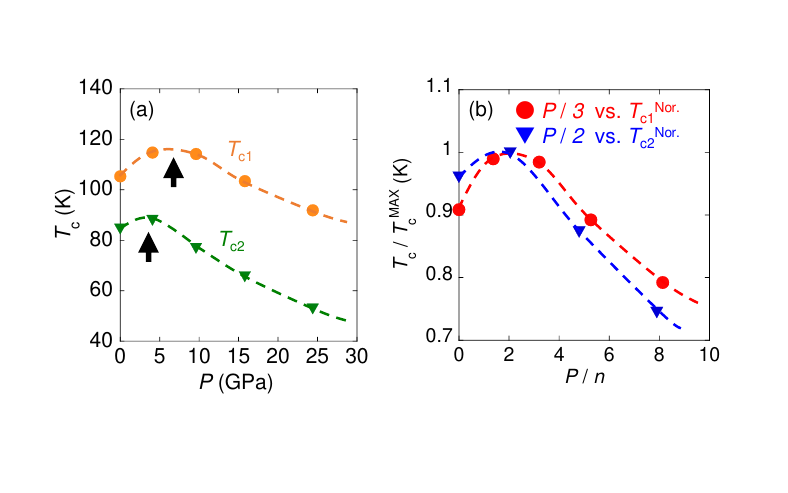}
\caption{(color online). (a) Pressure dependence of $\it T$$_{c1}$ and $\it T$$_{c2}$. The arrows indicate maximum $\it T$$_c$. (b) The dot circles show $\it P$ / 3 versus normalized $\it T$$_{c1}$. The inverse triangle dots show $\it P$ / 2 versus normalized $\it T$$_{c2}$.}
\label{f5}
\end{figure}

\acknowledgment
We are grateful to N. Sasaki, K. Kosugi, T. Usui, S. Iwasaki, M. Tanaka, S. Harada, T. Ishiyama, and T. Fujii for all the help. This work was supported by JSPS KAKENHI Grant No. JP17J05926, JST-Mirai Program Grant No. JPMJMI17A2, and JST CREST Grant No. JPMJCR16Q6.


\begin{thebibliography}{9}
\bibitem{Gao1994} L. Gao, Y. Y. Xue, F. Chen, Q. Xiong, R. L. Meng, D. Ramirez, and C. W. Chu, J. H. Eggert, and H. K. Mao, Phys. Rev. B $\bf 50$, R4260 (1994).
\bibitem{Monteverde2005} M. Monteverde, C. Acha, M. Nunez-Regueiro, D. A. Pavlov, K. A. Lokshin, S. N. Putilin, and E. V. Antipov, Europhys. Lett. $\bf 72$, 458 (2005).
\bibitem{Takeshita2013} N. Takeshita, A. Yamamoto, A. Iyo, and H. Eisaki, J. Phys. Soc. Jpn. $\bf 82$, 023711 (2013).
\bibitem{Mito2017} M. Mito, K. Ogata, H. Goto, K. Tsuruta, K. Nakamura, H. Deguchi, T. Horide, K. Matsumoto, T. Tajiri, H. Hara, T. Ozaki, H. Takeya, and Y. Takano, Phys. Rev. B $\bf 95$, 064503 (2017).
\bibitem{Mito2016} M. Mito, H. Goto, H. Matsui, H. Deguchi, K. Matsumoto, H. Hara, T. Ozaki, H. Takeya, and Y. Takano, J. Phys. Soc. Jpn. $\bf 85$, 024711 (2016).
\bibitem{Murayama1991} C. Murayama, Y. Iye, T. Enomoto, N. Mori, Y. Yamada, T. Matsumoto, Y. Kubo, Y. Shimakawa and T. Manako, Physica C $\bf 183$, 277 (1991).
\bibitem{Yoneda1990} T. Yoneda, Y. Mori, Y. Akahama, M. Kobayashi, and H. Kawamura, Jpn. J. Appl. Phys. $\bf 29$, 1396 (1990).
\bibitem{Watanabe2003} T. Watanabe, K. Tokiwa, M. Moriguch,  R. Horike, A. Iyo, Y. Tanaka, H. Ihara, M. Ohashi, M. Hedo, Y. Uwatoko, and N. Mori, J. Low Temp. Phys. $\bf 131$, 681 (2003).
\bibitem{Chen2010} X-J. Chen, V. V. Struzhkin, Y. Yu, A. F. Goncharov, C-T. Lin, H. K. Mao, and R. J. Hemley, Nature $\bf 466$, 950 (2010).
\bibitem{Akimitsu1987} J. Akimitsu, A. Yamazaki, H. Sawa, and H. Fujiki, Jpn. J. Appl. Phys. $\bf 26$, L2080 (1987).
\bibitem{Maeda1988} H. Maeda, Y. Tanaka, M. Fukutomi, and T. Asano, Jpn. J. Appl. Phys. $\bf 27$, L209 (1988).
\bibitem{Adachi1988} H. Adachi, S. Kohiki, K. Setsune, T. Mitsuyu, and K. Wasa, Jpn. J. Appl. Phys. $\bf 27$, L1883 (1988).
\bibitem{Parkin1988} S. S. P. Parkin, V. Y. Lee, E. M. Engler, A. I. Nazzal, T. C. Huang, G. Gorman, R. Savoy, and R. Beyers, Phys. Rev. Lett. $\bf 60$, 2539 (1988).
\bibitem{Schilling1993} A. Schilling, M. Cantoni, J. D. Guo, and H. R. Ott, Nature $\bf 363$, 56 (1993).
\bibitem{Scott1994} B. A. Scott, E. Y. Suard, C. C. Tsuei, D. B. Mitzi, T. R. McGuire, B. -H. Chen, and D. Walker, Physica C $\bf 230$, 239 (1994).
\bibitem{Karppinen1999} M. Karppinen and H. Yamauchi, Mater. Sci. Eng. $\bf 26$, 51 (1999).
\bibitem{Iyo2001} A. Iyo, Y. Aizawa, Y. Tanaka, M. Tokumoto, K. Tokiwa, T. Watanabe, and H. Ihara, Physica C $\bf 357$, 324 (2001).
\bibitem{Tokunaga1999} Y. Tokunaga, H. Kotegawa, K. Ishida, G-q. Zheng, Y. Kitaoka, K. Tokiwa, A. lyo, and H. Ihara, J. Low Temp. Phys. $\bf 117$, 437 (1999).
\bibitem{Kotegawa1999} H. Kotegawa, Y. Tokunaga, , K. Ishida, G.-q. Zheng, Y. Kitaoka, K. Tokiwa, A. Iyo, and H. Ihara, Phys. Rev. B $\bf 64$, 064515 (1999).
\bibitem{Mukuda2006} H. Mukuda, M. Abe, Y. Araki, Y. Kitaoka, K. Tokiwa, T. Watanabe, A. Iyo, H. Kito, and Y. Tanaka, Phys. Rev. Lett. $\bf 96$, 087001 (2006).
\bibitem{Shimizu2011} S. Shimizu, S-i. Tabata, H. Mukuda, and Y. Kitaoka, Phys. Rev. B $\bf 83$, 214514 (2011).
\bibitem{Mukuda2012} H. Mukuda, S. Shimizu, A. Iyo, and Y. Kitaoka, J. Phys. Soc. Jpn. $\bf 81$, 011008 (2012).
\bibitem{Iwai2014} S. Iwai, H. Mukuda, S. Shimizu, Y. Kitaoka, S. Ishida, A. Iyo, H. Eisaki, and S. Uchida, JPS Conf. Proc. $\bf 1$, 012105 (2014).
\bibitem{Adachi2015-2} S. Adachi, T. Usui, K. Takahashi, K. Kosugi, T. Watanabe, T. Nishizaki, T. Adachi, S. Kimura, K. Sato, K. M. Suzuki, M. Fujita, K. Yamada, and T. Fujii, Physics Procedia $\bf 65$, 53 (2015).
\bibitem{Fujii2001} T. Fujii, T. Watanabe, and A. Matsuda, J. Cryst. Growth $\bf 223$, 175 (2001).
\bibitem{Iye2010} T. Iye, T Nagatochi and A. Matsuda, Physica C $\bf 470$, 121 (2010).
\bibitem{Adachi2015} S. Adachi, T. Usui, Y. Ito, H. Kudo, H. Kushibiki, K. Murata, T. Watanabe, K. Kudo, T. Nishizaki, N. Kobayashi, S. Kimura, M. Fujita, K. Yamada, T. Noji, Y. Koike, and T. Fujii, J. Phys. Soc. Jpn. $\bf 84$, 024706 (2015).
\bibitem{Matsumoto2016} R. Matsumoto, Y. Sasama, M. Fujioka, T. Irifune, M. Tanaka, T. Yamaguchi, H. Takeya, and Y. Takano, Rev. Sci. Instrum., $\bf 87$, 076103 (2016).
\bibitem{Matsumoto2018} R. Matsumoto, A. Yamashita, H. Hara, T. Irifune, S. Adachi, H. Takeya, and Y. Takano, Appl. Phys. Express $\bf 11$, 053101 (2018).
\bibitem{Matsumoto2018-2} R. Matsumoto, H. Hara, H. Tanaka, K. Nakamura, N. Kataoka, S. Yamamoto, A. Yamashita, S. Adachi, T. Irifune, H. Takeya, and Y. Takano, J. Phys. Soc. Jpn. $\bf 87$, 124706 (2018).
\end{thebibliography}
\end{document}